\documentclass[11pt]{amsart}
\usepackage{amsmath}
\usepackage{amsxtra}
\usepackage{amscd}
\usepackage{amsthm}
\usepackage{amsfonts}
\usepackage{amssymb}
\usepackage{eucal}

\newcommand{\nn}{\nonumber}
\newcommand{\bea}{\begin{eqnarray}}
\newcommand{\ena}{\end{eqnarray}}
\newcommand{\be}{\begin{eqnarray*}}
\newcommand{\en}{\end{eqnarray*}}

\renewcommand{\ker}{\mathop{\rm Ker}}
\renewcommand{\Im}{\mathop{\rm Im}}
\newcommand{\End}{\mathop{{\rm End}}}
\newcommand{\ve}{\varepsilon}

\newcommand{\Z}{{\mathbb Z}}  
\newcommand{\C}{{\mathbb C}}  

\newcommand{\Q}{{\mathbb Q}}
\newcommand{\A}{\mathcal{A}}  
\newcommand{\W}{\overline{W}}  
\begin{document}

\title
{Central elements of the 
elliptic Yang--Baxter algebra at roots of unity}
\author{A.~Belavin and M.~Jimbo}
\address{AB: Landau institute for 
Theoretical Physics, Chernogolovka,
142432, Russia}\email{belavin@itp.ac.ru}  
`	\address{MJ: Graduate School of Mathematical Sciences, 
University of
Tokyo, Tokyo 153-8914, Japan}\email{jimbomic@ms.u-tokyo.ac.jp}

\date{hep-th/0208224}

\setcounter{footnote}{0}\renewcommand{\thefootnote}{\arabic{footnote}}

\begin{abstract}
We give central elements of the Yang-Baxter algebra 
for the $R$-matrix of the eight-vertex model, 
in the case when the crossing parameter 
is a rational multiple of one of the periods. 
\end{abstract}
\maketitle

\medskip

\noindent{\bf 1.}\quad 
As usual, let $R(u)$ be an $R$-matrix acting  
on  $\C^n\otimes \C^n$ 
and satisfy the Yang--Baxter equation 
\bea
&& R_{12}(u_1-u_2)R_{13}(u_1-u_3)R_{23}(u_2-u_3)
\nn\\
&&=R_{23}(u_2-u_3)R_{13}(u_1-u_3)R_{12}(u_1-u_2),
\label{YBE}
\ena
where $R_{12}(u)=R(u)\otimes I$, $R_{23}=I\otimes R(u)$, etc.. 
Here and after, we indicate by suffix the tensor components on which 
operators act nontrivially. 
By a {\it Yang-Baxter algebra} $\A$, 
we mean the algebra 
generated by the entries of an $n{\times}n$ matrix $L(u)$, 
subject to the defining relations 
\begin{equation}
R_{12}(u-v)L_1(u)L_2(v)=L_2(v)L_1(u)R_{12}(u-v),
\label{YBA}
\end{equation}
where $L_1(u)=L(u)\otimes I$ and $L_2(u)=I\otimes L(u)$.

The simplest and best-known solutions of \eqref{YBE} arise for $n=2$
and correspond to six- and eight-vertex models \cite{B}.
In the latter case, 
the elements of the $R$-matrix are written in terms of the
Jacobi elliptic theta functions with the half-periods $K$ and $K'$ as
\begin{equation}
\begin{aligned}
{}&R_{\ve\ve}^{\ve\ve}(u)=
\rho\Theta(\lambda)\Theta(\lambda u)H(\lambda(u+1)),
\\
{}&R_{\ve\ve'}^{\ve\ve'}(u)=
\rho\Theta(\lambda)H(\lambda u)\Theta(\lambda(u+1)),
\\
{}&R_{\ve'\ve'}^{\ve\ve}(u)=
\rho H(\lambda)H(\lambda u)H(\lambda(u+1)),
\\
{}&R_{\ve'\ve}^{\ve\ve'}(u)=
\rho H(\lambda)\Theta(\lambda u)\Theta(\lambda(u+1)),
\end{aligned}
\label{8vR}
\end{equation}
where $\ve,\ve'=\pm$ ($\ve \neq \ve'$)
are the labels of basis vectors $e_\pm$ of $V=\C^2$. 
The crossing parameter $\lambda$ plays an important role in the structure
of $\A$, its representations, and the integrable models connected with $\A$.
As can be seen from \eqref{YBA}, the parameter $\rho$ is inessential.

In this paper, we focus attention to the center of 
the Yang--Baxter algebra for the eight-vertex model 
when $p=\lambda/2K$ is a rational number.
We refer to this situation as the `root of unity' case.
We shall consider the case $p=2m/N$ 
where $m,N$ are coprime positive integers with 
$N\ge 3$ being odd. 

In the trigonometric degeneration $K'=0$,  
and for a general value of $\lambda$, 
it is well known that the center of $\A$ is generated by the 
quantum determinant. 
It is also well known that at 
roots of unity the center is enlarged. 
In this case, Tarasov \cite{T} gave an explicit expression
for the additional central elements. 
In terms of the entries $L_{\ve\ve'}(u)$ of $L(u)$ it reads 
\begin{equation}
\langle L_{\ve\ve'}\rangle(u)=
L_{\ve\ve'}(u+N-1)L_{\ve\ve'}(u+N-2)\cdots L_{\ve\ve'}(u).
\label{Tnew}
\end{equation}

Formula \eqref{Tnew} 
has the following elliptic counterpart.
Fix $u_0\in\C\backslash Z$ where $Z=\Q K+\Q K'$,  
and set for $a,b\in\Z$ 
\be
L_{\ve\ve';a,b}(u;u_0)=
\phi^*_{a, a+\ve}(u_0)L_{\ve\ve'}(u)\phi_{b, b+\ve'}(u_0), 
\en
where $\phi_{a,b}(u)$, $\phi^*_{a,b}(u)$ 
are Baxter's
intertwining vectors given in \eqref{int1}, \eqref{int2} below. 
Then the elements
\bea
&&\Lambda_{\ve\ve'}(u;u_0)
=\frac{1}{N}\sum_{a=0}^{N-1}
L_{\ve\ve';a,b}(u+N-1;u_0)L_{\ve\ve';a+\ve, b+\ve'}(u+N-2;u_0)
\cdots
\label{Bnew}\\
&&
\qquad\qquad\times
L_{\ve\ve';a+(N-1)\ve, b+(N-1)\ve'}(u;u_0)
\nn
\ena
do not depend on $b\in\Z$, and are central in $\A$. 
For different choices of the parameter $u_0$, 
the matrices 
$\Lambda(u;u_0)=\bigl(\Lambda_{\ve\ve'}(u;u_0)\bigr)$
and $\Lambda(u;u_0')$ are related by conjugation by 
a numerical matrix. 

Formula \eqref{Bnew} is a straightforward extension of 
Tarasov's formula \eqref{Tnew},
and we have no intention to claim any sort of originality.   
As we have not been able to find \eqref{Bnew}
in the literature, we give its derivation in this note.
\medskip

\noindent{\bf 2.}\quad 
%\section{}
From now on we specialize the crossing parameter as 
\bea
\lambda=\frac{4m}{N}K,
\label{lambda}
\ena
where $m,N$ are mutually 
prime positive integers and $N\ge 3$ is odd.
We fix a basis $e_+,e_-$ of $V=\C^2$ 
and $e_+^*,e_-^*$ of the dual space $V^*$ 
with the coupling $e^*_\ve e_{\ve'}=\delta_{\ve\ve'}$. 

The method of constructing central elements is 
a modification of the one in \cite{T}.
The outline is as follows. 
Let $V_i$ ($i=0,\cdots,N$) stand for copies of $V$. 
Define 
\begin{equation}
\begin{aligned}
{}&L_{1\dots N}(u)=L_1(u+N-1)\cdots L_N(u),
\\
{}&R_{1\dots N,0}(u)=R_{10}(u+N-1)\cdots R_{N0}(u).
\end{aligned}
\label{Ldef}
\end{equation}
Both $L_{1\dots N}(u)$ and $R_{1\dots N,0}(u)$ 
are operators on $V_1\otimes\cdots\otimes V_N$, 
whose entries belong to $\A$ and 
$\End(V_0)$ respectively.  
From the form of the $R$-matrix \eqref{8vR}, together with 
\eqref{YBE} and \eqref{YBA}, 
it follows for any value of $\lambda$ that 
$R_{1\dots N,0}(u)$ and $L_{1\dots N}(u)$ 
leave invariant a subspace 
$W_0\subset V_1\otimes \cdots\otimes V_N$ 
of codimension $N+1$. 
Moreover, if $\lambda$ satisfies \eqref{lambda}, 
then $W_0$ is contained in a larger invariant subspace $W$,
such that the quotient space $\W=W/W_0$ is    
$2$-dimensional. 

Consider the restriction of 
$L_{1\dots N}(u)$ and $R_{1\dots N,0}(u)$ 
to $\W$ and denote the resulting operators 
by $\Lambda(u)\in\End(\W)\otimes\A$  
and $\overline R(u)\in\End(\W)\otimes \End(V_0)$. 
We have
\begin{equation}
\overline R(u-v) \Lambda(u)L_0(v)=L_0(v)\Lambda(u)\overline R(u-v).
\label{YBB}
\end{equation}
We then use the results in \cite{D} to show that 
\begin{equation}
\overline R(u)=f(u)I\otimes I,
\label{hatR}
\end{equation}
where $f(u)$ is a non-vanishing scalar factor.
From \eqref{YBB}, we therefore have
\begin{equation}
[\Lambda(u),L_0(v)]=0, 
\label{LamL}
\end{equation}
which shows that 
the entries of $\Lambda(u)$ are central elements of $\A$. 
The space $\W$ has a basis
$\{\Psi_\ve(u_0)\}_{\ve=\pm}$,  
which are obtained by $N$-fold 
fusion of the intertwining vectors. 
Writing $\Lambda(u)$ in this basis we arrive at \eqref{Bnew}. 
\medskip

\noindent{\bf 3.}\quad 
In this and the next subsection,  
we recall some aspects of the standard fusion procedure. 

{}From \eqref{YBE} and \eqref{YBA}, we have
\begin{equation}
\begin{aligned}
{}&R_{i,i+1}(1)R_{1\dots N,0}(u)= R_{1\dots N,0}^{i,i+1}(u)R_{i,i+1}(1),
\\
{}&R_{i,i+1}(1)L_{1\dots N}(u)= L_{1\dots N}^{i,i+1}(u)R_{i,i+1}(1),
\end{aligned}
\label{RR}
\end{equation}
with some operators $R_{1\dots N,0}^{i,i+1}(u)$ 
and $L_{1\dots N}^{i,i+1}(u)$ 
which differ from $R_{1\dots N,0}(u)$ and $L_{1\dots N}(u)$ by
permutation of the $i$th and $(i{+}1)$th factors. 
Taking $u=1$ in \eqref{8vR}, we see that $R_{i,i+1}(1)$
has a kernel spanned by the vectors
$$
e_{\alpha_1}\otimes\cdots\otimes e_{\alpha_{i-1}}\otimes
(e_{+}\otimes e_{-}-e_{-}\otimes e_{+})\otimes e_{\alpha_{i+1}}
\otimes \cdots \otimes e_{\alpha_N},
$$
where $\alpha_k=\pm1$.
The subspace
\bea
&&W_0=\sum_{i=1}^{N-1}\ker R_{i,i+1}(1)
\quad \subset V_1\otimes\cdots\otimes V_N 
\label{W_0def}
\ena
is invariant 
under the action of $R_{1\dots N,0}(u)$ and $L_{1\dots N}(u)$.
The subspace of symmetric tensors
in $V_1\otimes\cdots\otimes V_N$
is isomorphic to  
$V_1\otimes\cdots\otimes V_N/W_0$.  
We have $P v\equiv v\bmod W_0$ ($v\in V_1\otimes\cdots\otimes V_N$) 
for any permutation $P$ of the tensor components. 

Introduce the operator 
\begin{equation}
S=S^{(1)}S^{(2)}\cdots S^{(N-1)},
\quad S^{(j)}=R_{j\,j+1}(j)\cdots R_{23}(2)R_{12}(1).
\label{Sdef}
\end{equation}
It follows from \eqref{YBE} and \eqref{YBA} that
\begin{equation}
\begin{aligned}
{}&SR_{1\dots N,0}(u)= R_{N\dots 1,0}(u)S,
\\
{}&SL_{1\dots N}(u)= L_{N\dots 1}(u)S,
\end{aligned}
\label{SR}
\end{equation}
where $R_{N\dots 1,0}(u)$ and $L_{N\dots 1}(u)$ differ from
$R_{1\dots N,0}(u)$ and $L_{1\dots N}(u)$ 
by permutation of all factors to the opposite order.
Therefore
\bea
&&W=\ker S
\label{Wdef}
\ena
is also invariant 
under $R_{1\dots N,0}(u)$ and $L_{1\dots N}(u)$.
%For instance, if $N=3$ and $i=2$, then 
%$S=R_{12}(1) R_{13}(2) R_{23}(1)$.
By using \eqref{YBE}, for any given $i$ 
the factors of $S$ can be reordered
in such a way that $R_{i,i+1}(1)$ comes to
the rightmost position. 
Hence we have $W_0\subset W$. 
\medskip

\noindent{\bf 4.}\quad 
Let us show that $\W=W/W_0$ is $2$-dimensional
if the elliptic modulus is generic.  

In the trigonometric limit, it is verified as follows.  
With an appropriate base change, 
the operator $S$ commutes with an 
action of $U_q(\frak{sl}_2)$ where $q$ is a 
primitive $N$-th root of unity. One checks that 
$\Im S$ is a proper nontrivial submodule of 
the specialization of the standard $N+1$ dimensional representation.
From the representation theory of $U_q(\frak{sl}_2)$ at $q^N=1$, we conclude
that $\dim \Im S=N-1$, and hence $\dim \ker S/W_0=2$. 

From the above consideration, we have 
$\dim W/W_0\le 2$ in the elliptic case 
if the modulus is generic.
We show the equality by constructing two linearly independent 
elements of $\W$.
For this purpose we use Baxter's intertwining vectors 
$\phi_{a,b}(u)\in V$ and their duals
$\phi^*_{a,b}(u)\in V^*$ 
given in \eqref{int1},\eqref{int2}.  
Their main properties are summarized in the appendix. 

Set 
\bea
&&\Psi_{\ve,a}(u)=
\phi_{a,a+\ve}(u+N-1)\otimes \phi_{a+\ve,a+2\ve}(u+N-2)\otimes 
\cdots \otimes
\phi_{a+(N-1)\ve,a+N\ve}(u),
\label{Psidef}
\\
&&
\Psi^*_{\ve,a}(u)=
\phi^*_{a,a+\ve}(u+N-1)\otimes \phi^*_{a+\ve,a+2\ve}(u+N-2)\otimes 
\cdots \otimes
\phi^*_{a+(N-1)\ve,a+N\ve}(u).
\label{Psi*def}
\ena

We will use the following properties:
\begin{enumerate}
\item $S\Psi_{\ve,a}(u)=0$. 
\item $\Psi^*_{\ve,a}(u)\Psi_{\ve',b}(u)=\delta_{\ve,\ve'}$
\quad ($a,b\in\Z, \ve,\ve'=\pm$).
\item 
\bea
\Psi_{\ve,a}(u)\equiv \Psi_{\ve,b}(u)
~~\bmod W_0
\quad (a,b\in\Z, \ve,\ve'=\pm).
\label{indep}
\ena
\item 
The vector
\be
&&\Psi^*_{\ve}(u)=\frac{1}{N}\sum_{a=0}^{N-1}\Psi^*_{\ve,a}(u)
\label{Psi*}
\en
is orthogonal to $W_0$. 
\end{enumerate}
Assertion (i) is a consequence of 
the vertex-face correspondence \eqref{VF}.  
The orthogonality (ii) 
follows from \eqref{ortho1}--\eqref{ortho2}.
To see (iii), we use the expression \eqref{int1}, 
$\phi_{a,a+\ve}(u)=\phi(a-\ve u)$, $\phi(u)=\phi(u+N)$,    
to find
\be
\Psi_{\ve,a+2\ve}(u)=C\Psi_{\ve,a}(u)
\equiv \Psi_{\ve,a}(u)~~\bmod W_0,
\en 
where $C$ is a cyclic permutation of the tensor components.  
Since $N$ is odd, (iii) follows. 
One can verify (iv) in a similar manner.

From (ii) and (iv), it is clear that 
the two vectors in $\W=W/W_0$ 
\be
\Psi_{\ve}(u)=\Psi_{\ve,a}(u)~~\bmod W_0
\qquad (\ve=\pm)
\en
are linearly independent for any $u\in\C\backslash Z$. 
\medskip

\noindent{\bf 5.}\quad
Let $\overline R(u)$ and $\Lambda(u)$ be restrictions 
of $R_{1\dots N,0}(u)$ and $L_{1\dots N}(u)$ on $\W$.
Fixing $u_0\in \C\backslash Z$, we have in 
the basis $\{\Psi_{\ve}(u_0)\}_{\ve=\pm}$
\be
\begin{aligned}
{}&L_{1\dots N}(u) \Psi_{\ve'}(u_0)=
\sum_{\ve=\pm}\Psi_\ve(u_0)
\Lambda_{\ve\ve'}(u;u_0)
\bmod W_0,
\end{aligned}
\label{hRdef}
\en
where
\be
\Lambda_{\ve\ve'}(u;u_0)
=\Psi^*_\ve(u_0)L_{1\dots N}(u) \Psi_{\ve'}(u_0).
\en
The right hand side yields the formula \eqref{Bnew}. 

To complete the proof of the centrality \eqref{LamL}, 
it remains to verify \eqref{hatR}.
{}For this, it suffices to show the relation
\bea
R_{1\dots N,0}(u-v) \Psi_{\ve,a}(u)\otimes \phi_{a,a+\ve'}(v)=
f(u-v) \Psi_{\ve,a}(u)\otimes \phi_{a,a+\ve'}(v) \bmod W_0.
\label{R=1}
\ena

In the present notation, 
the fused intertwining vectors $\phi_{N,a,a+N\ve}(u)$ 
of (2.3.8) in \cite{D} is $P_{1\dots N}\Psi_{\ve,a}$,  
where $P_{1\dots N}$ denotes the complete symmetrizer 
in $V_1\otimes\cdots\otimes V_N$. 
Noting 
$R_{1\dots N,0}(u)P_{1\cdots N}=R_{1\cdots N,0}(u)$, 
we have from Theorem 2.3.3 in \cite{D}
\be
&&R_{1\dots N,0}(u-v) \Psi_{\ve,a}(u)\otimes \phi_{a,a+\ve'}(v)
\nonumber
\\
&&=P_{1\dots N}R_{1\dots N,0}(u-v)
\phi_{N,a,a+N\ve}(u) \otimes \phi_{a,a+\ve'}(v)
\nonumber
\\
&&=\tilde f(u-v) \sum_{\ve''=\pm} 
\mathcal{W}_{N1}(a+\ve'',a+\ve',a,a|u-v)
\nonumber
\\
&&
\qquad\qquad\qquad
\times
\phi_{N,a-(N-1)\ve'+\ve'',a+\ve'}(u)\otimes
\phi_{a,a+\ve''}(v) \bmod W_0,
\en
where $\tilde f(u)$ is a non-vanishing scalar function 
(see the end of p.45, \cite{D}), 
and $\mathcal{W}_{N1}(a,b,c,d|u)$ are the Boltzmann 
weights of the fusion SOS models. 
We have used its periodicity by $N$ with respect to $a,b,c,d$. 
From the explicit formula (2.1.20) in~\cite{D} we have 
\be
\mathcal{W}_{N1}(a+\ve'',a+\ve',a,a|u)
=
\delta_{\ve\ve''}g(u),
% \frac{H(\lambda u)\Theta (\lambda u)}{H(\lambda )\Theta(\lambda)}.
\en
where $g(u)$ is another non-zero scalar. 

Returning to the notation $\Psi_{\ve,a}(u)$ and setting $f(u)=\tilde f(u) g(u)$, 
we find
\be
R_{1\dots N,0}(u-v) \Psi_{\ve,a}(u)\otimes \phi_{a,a+\ve'}(v)=
f(u-v) \Psi_{\ve',a}(v)\otimes \phi_{a,a+\ve'}(v) \bmod W_0.
\label{hatR=1}
\en
In view of \eqref{indep}, we arrive at \eqref{R=1}.	
\bigskip

{\bf Acknowledgments.}
One of the authors (A.~B.) 
thanks the Graduate School of Mathematical 
Sciences, The University of Tokyo, 
where this work was done, for hospitality.
This work is supported by RFBR-01-02-16686,00-15-96579;
INTAS-00-00055;CRDF-RP1-2254 (A.~B.).
\bigskip

\noindent{\bf Appendix.}\quad
Here we collect known formulas concerning intertwining vectors. 

Following the notation of \cite{D}, pp.46--47,  
we choose $s^+=s^-=\xi+K/\lambda$, 
where $\lambda$ is specified as in \eqref{lambda}, and 
$\xi$ is a generic complex parameter. 
For $a\in\Z$ and $\ve=\pm 1$, Baxter's intertwining vectors 
are given by 
\bea
&&\phi_{a,a+\ve}(u)=\phi(a-\ve u),
\nn\\
&&\phi(u)=H_1\bigl(\lambda(\xi+ u)\bigr)e_+
+\Theta_1\bigl(\lambda(\xi+ u)\bigr)e_-,
\label{int1}
\ena
where $H_1(u)=H(u+K),\Theta_1(u)=\Theta(u+K)$. 
The dual intertwining vectors are defined by 
$\phi^*_{a+\ve,a}(u)\phi_{a+\ve',a}(u)=\delta_{\ve\ve'}$. 
Explicitly we have 
\bea
&&\phi^*_{a+\ve,a}(u)=\frac{\ve}{\Delta_a(u)}\phi^*(a-\ve-\ve u),
\nn\\
&&\phi^*(u)=-\Theta_1\bigl(\lambda(\xi+ u)\bigr)e_+^*
+H_1\bigl(\lambda(\xi+ u)\bigr)e_-^*.
\label{int2}
\ena
Here 
\be
&&\Delta_a(u)=[1+u]
\frac{[\xi+a+\frac{2K}{\lambda}]}{\zeta [\frac{K}{\lambda}]}, 
\\
&&
[u]=H(\lambda u)\Theta(\lambda u),
\en
and $\zeta$ is a non-zero complex number depending 
only on $K'/K$. Notice that $[u+N]=[u]$, $[N]=0$. 
We have 
\be
\phi(u+N)=\phi(u),\quad 
\phi^*(u+N)=\phi^*(u).
\en
We set $\phi_{a,b}(u)=\phi^*_{a,b}(u)=0$ unless $b=a\pm 1\bmod N$. 
In other words, we consider a cyclic SOS model
rather than a restricted SOS model. 

We have the vertex-face correspondence 
\begin{equation}
R(u-v)\phi_{d,c}(u)\otimes \phi_{c,b}(v)=
\sum_{d}\mathcal{W}_{11}(a,b,c,d|u-v)
\phi_{a,b}(u)\otimes \phi_{d,a}(v),
\label{VF}
\end{equation}
where 
$\mathcal{W}_{11}(a,b,c,d|u)$ denotes 
the Boltzmann weights of the SOS model
as given in (2.1.4a)--(2.1.4c), \cite{D}.

We have also the relations
\bea
&&\phi^*_{a\pm\ve, a}(u)\phi_{b\pm\ve, b}(u)
=\frac{[\xi+\frac{a+b}{2}]}{[\xi+a]}\frac{[u+1\mp\frac{a-b}{2}]}{[u+1]},
\label{ortho1}
\\
&&
\phi^*_{a\mp\ve,a}(u)\phi_{b\pm\ve, b}(u)
=\frac{[\frac{a-b}{2}]}{[\xi+a]}\frac{[u+1\pm(\xi+\frac{a+b}{2})]}{[u+1]}.
\label{ortho2}
\ena

\end{document}